\documentclass[runningheads,a4paper]{llncs}

\usepackage{amssymb}
\usepackage{graphicx}
\usepackage[numbers]{natbib}
\usepackage{amssymb}
\usepackage{array}
\usepackage{mdwmath}
\usepackage{mdwtab}
\usepackage{eqparbox}
\usepackage{subfig}
\usepackage{amsmath}

\usepackage{url}
\urldef{\mailsa}\path|{silvia.puglisi}@upc.edu|
\urldef{\mailsb}\path|{david.rebollo,jforne}@entel.upc.edu|    

\newcommand{\keywords}[1]{\par\addvspace\baselineskip
\noindent\keywordname\enspace\ignorespaces#1}

\begin{document}

\mainmatter  

\title{You never surf alone. Ubiquitous tracking of users' browsing habits.}

\titlerunning{Lecture Notes in Computer Science: Authors' Instructions}

%
%
\author{Silvia Puglisi%
\and David Rebollo-Monedero\and Jordi Forn\'e}
\authorrunning{Lecture Notes in Computer Science: Authors' Instructions}

\institute{{Department of Telematics Engineering, \\ 
Universitat Polit\`ecnica de Catalunya (UPC)\\
C.\ Jordi Girona 1-3, 08034 Barcelona, Spain}\\
\mailsa\\
\mailsb\\
\url{http://www.upc.edu}}

%
%

\toctitle{Lecture Notes in Computer Science}
\tocauthor{Authors' Instructions}
\maketitle

\begin{abstract}

In the early age of the internet users enjoyed a large level of anonymity. At the time web pages were just hypertext documents; almost no personalisation of the user experience was offered. The Web today has evolved as a world wide distributed system following specific architectural paradigms. On the web now, an enormous quantity of user generated data is shared and consumed by a network of applications and services, reasoning upon users expressed preferences and their social and physical connections. Advertising networks follow users' browsing habits while they surf the web, continuously collecting their traces and surfing patterns. We analyse how users tracking happens on the web by measuring their online footprint and estimating how quickly advertising networks are able to profile users by their browsing habits.

\end{abstract}

\section{Introduction}

When users surf the web an intricate network of \emph{personalisation services} tracks their preferences by \emph{following} their browsing habits. These data is used to provide tailored suggestions, in terms of products users could buy, resources they might find interesting, social connections they might be interest to form.
Personalisation services sometimes rely on different techniques to track users across different websites and applications. Many of these techniques use cookies. For example, Google Analytics service~\cite{GoogleAnalytics} uses cookies to measure user-interactions on websites.  Another set of these techniques uses web or app sessions left open by the user. As an example someone might decide to check their web email account and then continue to surf the web without signing out, therefore leaving their session open. 
Yet another set of these techniques uses personalised features of the user's device or browser to restrict the pool of possible candidates among their visitors. Features that might be used by advertising networks include personalised language or fonts settings, browser extensions and so on.
By identifying user through their accounts or unique features, analytics technologies can distinguish unique users across multiple devices or sessions.

\subsection{Contribution}

We have observed how users are tracked across the web and how the displayed advertising is tailored even after they have visited a few websites with a certain interest bias. In our study we analyse how quickly advertising networks can identify a user and start tracking them by measuring the distance between the measured user profile and the advertising profile.
We introduce a set of metrics to express this distance and measure the number of web sites visited after which the distance between the advertising profile and the user profile is less then a certain threshold.
\\
It is important to know that we have consider the case for which users are not registering, neither connecting any external account, as it could be the case with services like: Facebook, Google+, Twitter, and so on.
We shall also point out that we have concentrated our study onto a single advertising network: Google. We reserve to future studies the possibility to include and analyse also other networks.
\\
\noindent
The main contributions of this paper are the following.
\begin{enumerate}
 \item  Introducing a model of the user online footprint.
 \item Measuring how quickly a user is uniquely identified and tracked by an advertising network.
 \item Introducing a measure of similarity between the user profile and the observed advertising profile.
 \end{enumerate}

\section{Background}

Information regarding locations, browsing habits, communication records, health information, financial information, and general preferences regarding user online and offline activities are shared by different parties online. This level of access is often directly granted from the user of such services. In a wide number of occasion though, private information are captured by online services without the direct user consent or even knowledge. We believe that the privacy and sensitiveness of the information becoming accessible to third parties can be easily overlooked. 
\\
Personal computers and more generally communication devices that are carried around by people are capable of being located, identified and tracked across different locations, networks and services~\cite{michael2013location}. All these devices can therefore be used for a variety of surveillance activities, which are in itself detrimental to the user's interests. Until recently in fact, the cost of surveillance and tracking of people and activities was proportional to the cost of directly reaching, asking or following a single person or a group of people. Technology therefore enhances the surveillance capabilities by introducing tools that allow the collection of information arising from a person's activities. This information can furthermore be combined and inferred, therefore offering a more complete picture of that person. 
\\
For example, to personalise their services or offer tailored advertising, web applications could use tracking services that identify a user through different networks~\cite{veeningen2014line}~\cite{getoor2012entity}. These tracking services usually combine information from different profiles that users create, for example their Gmail address or their Facebook or LinkedIn accounts. In addition specific characteristics of the user's device can be used to identify them through different sessions and websites, as described by the Panopticlick project~\cite{eckersley2011panopticlick}.
\\
Browser fingerprinting is a technique implemented by analytics services and tracking technologies to identify uniquely a user while they browser different websites. Different features of a specific browser setup can be used to identify uniquely a user. Supported languages, browser extensions or installed fonts~\cite{boda2012user} can be used to identify a browser setup among others. More advanced techniques distinguish between browsers' JavaScript execution characteristics~\cite{mowery2011fingerprinting}. These features are particularly interesting since they are more difficult to simulate or mitigate in practice. Targeting JavaScript execution characteristics actually means looking at the innate performance signature of each browser's JavaScript engine, allowing the detection of browser version, operating system and microarchitecture. These attacks can also work in situations where traditional forms of system identification (such as the user-agent header) are modified or hidden. Other techniques exploit the whitelist mechanism of the popular NoScript Firefox extension.This mechanism allow the user to selectively enabling web pages' scripting privileges to increase privacy by allowing a site to determine if particular domains exist in a user's NoScript whitelist.
\\
It is important to note that while tracking creates serious privacy concerns for internet users, the customisation of results is also beneficial to the end user~\cite{castelluccia2012behavioural}. In fact, while tailored services offer to the user only information relevant to their interests, it also allows some companies and institutions to concentrate an enormous amount of information about internet users in general. ~\cite{rao2015they} investigate user profiling and access mechanisms offered by online data aggregator to users' collected data. Both the collected data and its accuracy was analysed together with the user's concerns. In their findings about 70\% of the participants to the study expressed some concerns about the collection of sensitive data, its level of detail and how it might be used by third parties, especially for credit and health information.
\\
It has been shown how most successful tracking networks exhibit a consistent structure across markets, with a dominant connected component that, on average, includes 92.8\% of network vertices and 99.8\% of the connecting edges~\cite{gomer2013network}. ~\cite{gomer2013network} have measured the chance that a user will become tracked by all top 10 trackers in approximately 30 clicks on search results to be of 99.5\%. More interesting, ~\cite{gomer2013network} have shown how tracking networks present properties of the small world networks. Therefore implying a high-level global and local efficiency in spreading the user information and delivering targeted ads.

\section{Modelling the user's footprint}

\begin{figure}[!ht]
\centering
\includegraphics[scale=0.45]{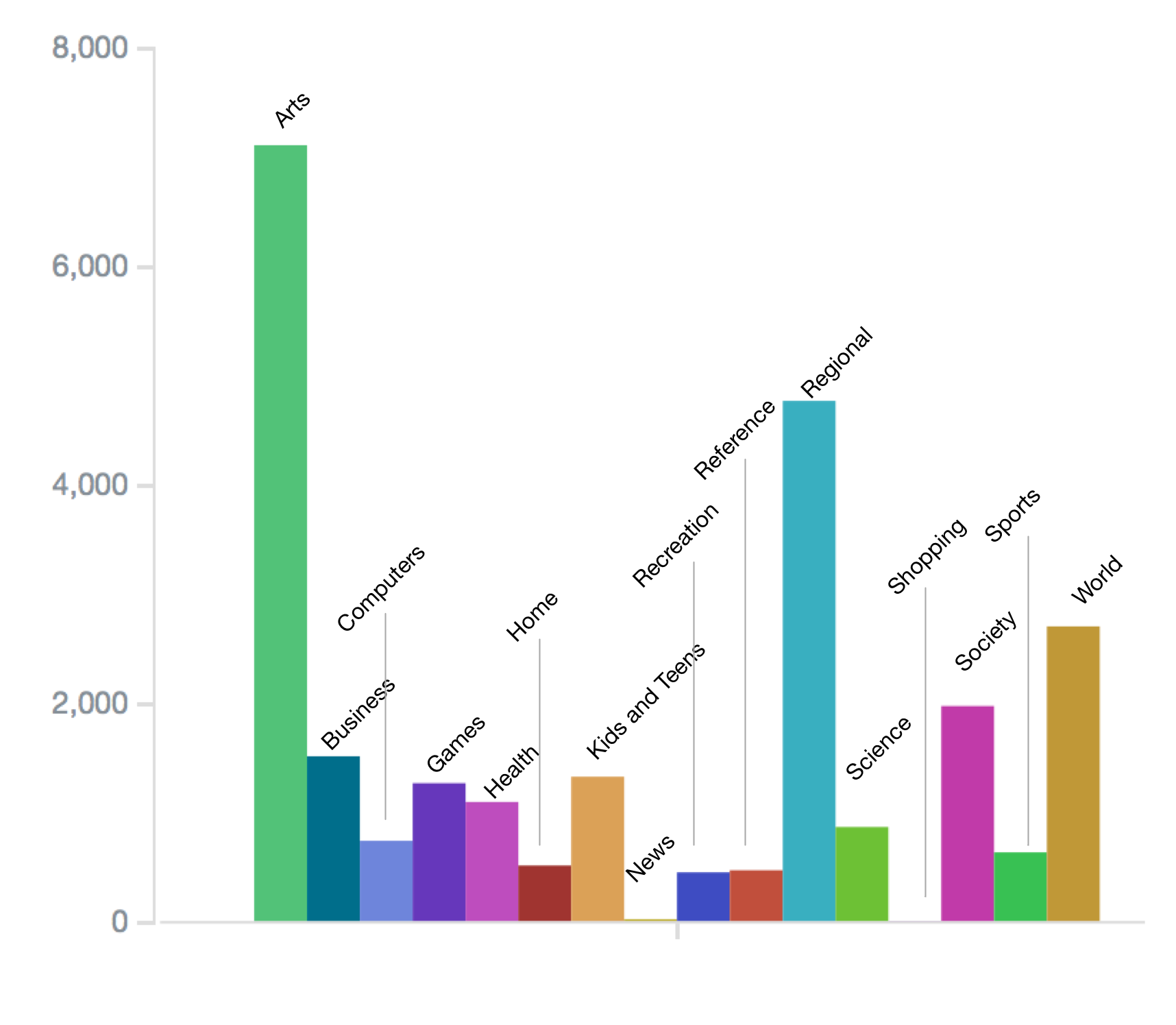}
\caption{Here we show an example of user profile expressed in absolute terms by counting the number of keywords in each category for a browsing session. We model user and advertising profiles as histograms of tags keywords a set of predefined categories of interest.
\label{abs-profile}}
\end{figure}

We model the user's activity as series of events belonging to a certain identity. Each event is a document containing different information. We can formally defined this as a hypermedia document i.e. an object possibly containing graphics, audio, video, plain text and hyperlinks. We call the hyperlinks selectors and we use these to build the connections between the user's different identities or events. Each identity is a profile that the user has created onto a service or platform. This can be an application account or a social network account, such as their LinkedIn or Facebook unique IDs. An event is an action performed by the user, like visiting a website or creating a post on a blog.
\\
We aggregate keywords each time the user creates a new event by visiting a different url. These keywords constitute the user profile of interests (Figure ~\ref{abs-profile}). A tractable model of the user profile as a probability mass function (PMF) is proposed in~\cite{Parra12DKE,Parra12TKDE} to express how each keyword contributes to expose how many times the user has indirectly expressed a preference toward a specific category. We consider that the user expresses a preference when they visit a webpage categorised with a certain keywords. This model follows the intuitive assumption that a particular category is weighted according to the number of times this has been counted in the user  profile.
\\
We define the profile of a user $u_m$ as the PMF $p_m = (p_{m,1},\ldots, p_{m,L})$, conceptually a histogram of relative frequencies of tags across the set of tag categories $\mathcal{T}$.
\\
Similarly, we define the profile of an ads $i_n$ as the PMF $q_n =(q_{n,1},\ldots, q_{n,L})$, where $q_{n,l}$ is the percentage of tags belonging to the category $l$ which have been assigned to this specific advertising item. Both user and ads profiles can then be seen as normalised histograms of tags across categories of interest. Our profile model is in this extent equivalent to the tag clouds that numerous collaborative tagging services use to visualise which tags are being posted, collaboratively or individually by each user. A tag cloud, similarly to a histogram, is a visual representation in which tags are weighted according to their relevance.
\\
In view of the assumptions described in the previous section, our privacy attacker boils down to an entity that aims to profile users by representing their interests in the form of normalised histograms, on the basis of a given categorisation.

\subsection{A metric of similarity}

We consider the third party advertising network to operate like a recommendation system, that suggest products or services that might be of interest for the user, based on their preferences. A recommendation system can be described as an information filtering system that seeks to predict if the user is interested or not in a particular resource. We assume that the ad server suggest advertising based on a measure of \emph{similarity}. 
\\
We measure the user profile, as previously described, as an histogram of their recorded preferences, and the advertising profile as an histogram of the ads that the user has received.  We use the $1-norm$ as a measurement of how the advertising network is tracking the user profile:

$$ \| p_{m}, q_{n} \|_1 = \sum_i{ | {p_m}_i - {q_n}_i | } $$

\section{Experimental methodology and results}

We analysed the browsing habits of 86 users of Twitter, by observing the set of websites they share in their feed. We assumed that the articles shared on twitter are a subset of the website that each users visit every day. Yet if they are active Twitter users, these websites will express their interest bias towards certain categories. Please note that we only consider the links shared on the platform as a sequence of website that the user might have visited. These sites are therefore surfed in our simulation environment. Here we pretend that a user is going through their reading list of sites and we measure how the advertising changes in the page and adapts to their profile. The user is simulated by a software agent opening the urls and surfing the page for a certain arbitrary amount of time.
\\
In our simulated environment the users are not logged in Twitter or any other account. 
\\
For each users we analysed the websites and collected keywords for the shared articles. We used both the meta information contained in the page, as well as extracted keywords from the actual text of the page by using the Rapid Automatic Keyword Extraction (RAKE)~\cite{RAKE} algorithm. Each keyword was evaluated against Open Directory Project (DMOZ)~\cite{a22} for classification within top levels categories.
\\
Once the user profile was calculated the advertising profile is evaluated. The advertising profile is extracted from url parameters contained in third party requests. We have considered only Google ads for the purpose of this study. These parameters are again evaluated against DMOZ for classification within top levels categories.
\\
At each step the \emph{linear norm} between the advertising profile and the actual user profile is evaluated. 

\begin{figure}[!ht]
\centering
\includegraphics[scale=0.55]{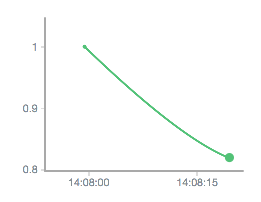}
\caption{The figure illustrates how the norm between the advertising and the browsing profile decrease of approximately 20\% in two subsequents visits and only in 15 seconds.
\label{1-norm}}
\end{figure}

\begin{figure}[!ht]
\centering
\includegraphics[scale=0.35]{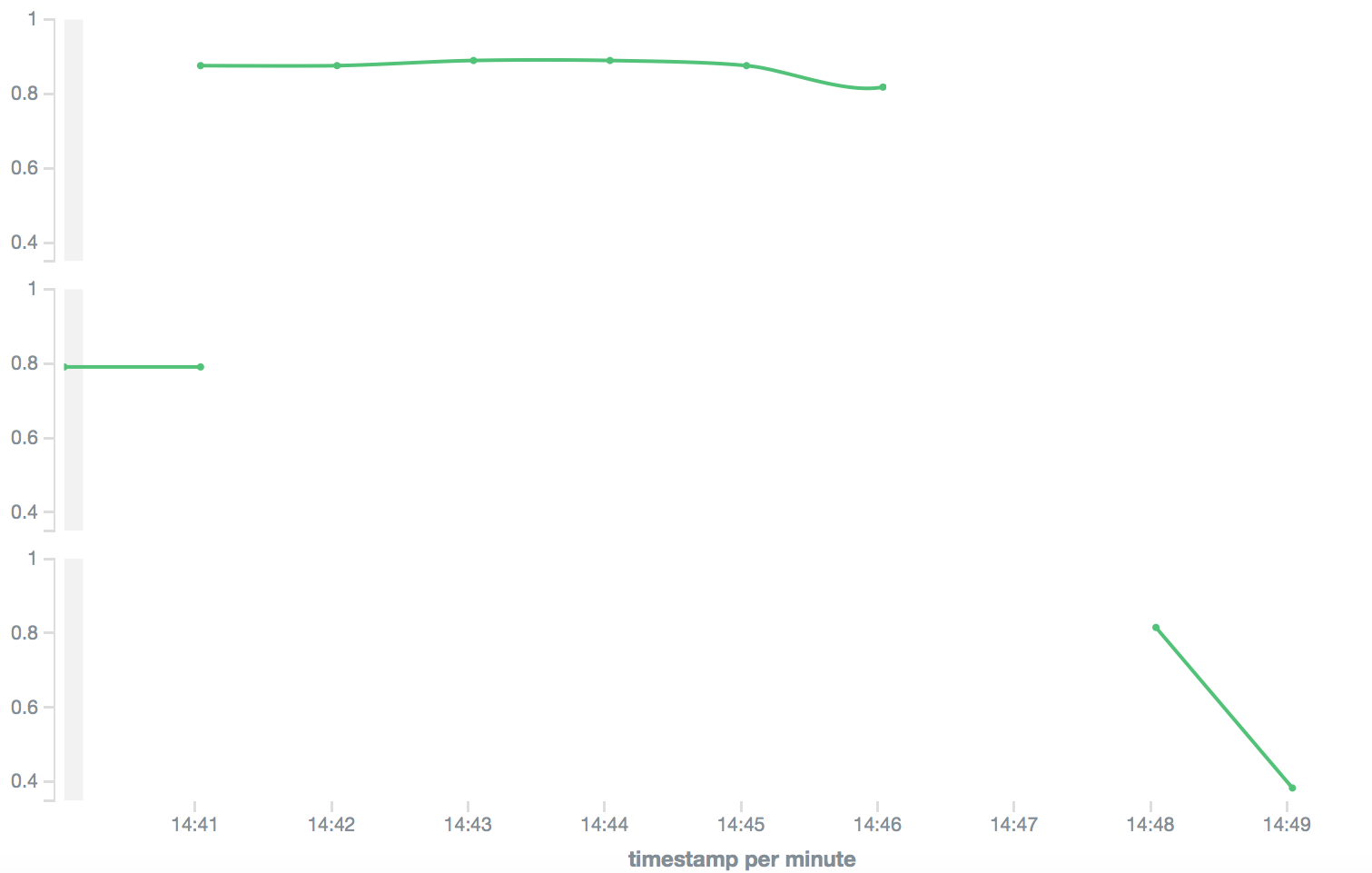}
\caption{The figure compare the 1-norm decrease for three different users in a short timespan. This shows how, when a user surf websites in a specific category the advertising slowly adapts to the new category and the norm decreases. When the category changes the advertising needs to adjust again. 
\label{users-1-norm}}
\end{figure}

By profiling users' browsing events using a hypermedia document structure we were able to show how each event contains a set of features regarding the user identity and the page that was visited. We have therefore categorised each event by using the keywords contained in the meta information present in the page and the page text itself (Figure ~\ref{abs-profile}). We have observed how a large and sophisticated advertising network such as Google is able to profile users quickly and only in a few visits (Figures ~\ref{1-norm} and ~\ref{users-1-norm}) 

We have found how our hypermedia model is particularly suited for big data analysis and how it allows a user to keep their online footprint under control by understanding precisely which websites have introduced certain keywords in their profile. Eventually this technique would allow the user to implement more precise Privacy Enhancing Techniques (PETs) to masquerade their profile to advertising networks.

\section{Conclusions and future work}

Using a hypermedia model to capture the footprint that users leave online while surfing the web has proven a promising technique. Particularly the model is able to capture both the single categorisation that each webpage introduce as well as time series analytics and breaking up of third party tracking per advertising network. 
We have also shown how web tracking by large advertising networks happens very quickly in a few subsequent visits to websites in the network (Figures ~\ref{1-norm} and ~\ref{users-1-norm}).
In future research we would like to further explore the hypermedia model introduced, while continuing to understand how quickly web advertising is able to match the served ads with the actual user profile. This would allow us to understand if different profiles for the same users can be somehow linked together within similar advertising networks.
We are also particularly interested in measuring how social networks sharing buttons and/or commenting services, included on websites, are able to track users even when these have not signed in with their account. We reserve the study of their capabilities to future investigations.
More over we want to enlarge the set of users analysed by testing on logs from a real world small computer network, while also introducing new metrics to our study. In particular we are already planning to consider: 2-norm, KL-Divergence between the advertising profile and the observed user profile, Fisher information.
We also believe in the importance to provide users with simple visualisation tools able to show the user their online footprint and allowing them to take action to masquerade their interests profile or simply block certain networks.

\subsection*{Acknowledgments.} This work was partly supported by the Spanish Government through projects CONSEQUENCE (TEC2010-20572-C02-02) and EMRISCO (TEC2013-47665-C4-1-R).

\bibliographystyle{splncs03}

\bibliography{Bibliography/StringAbbreviated,Bibliography/Security,Bibliography/InfoTheory,Bibliography/LosslessCoding,Bibliography/LossyCoding,Bibliography/MathStatSigPro,Bibliography/Classification,Bibliography/Applications,Bibliography/ReferencesTRIPP,Bibliography/SemanticWeb,Bibliography/InsubriaReferences,Bibliography/rfc,Bibliography/Silvia_bibliography}

\begin{thebibliography}{10}
\providecommand{\url}[1]{\texttt{#1}}
\providecommand{\urlprefix}{URL }

\bibitem{a22}
Open directory project - http://www.dmoz.com

\bibitem{boda2012user}
Boda, K., F{\"o}ldes, {\'A}.M., Guly{\'a}s, G.G., Imre, S.: User tracking on
  the web via cross-browser fingerprinting. In: Information Security Technology
  for Applications, pp. 31--46. Springer (2012)

\bibitem{castelluccia2012behavioural}
Castelluccia, C.: Behavioural tracking on the internet: a technical
  perspective. In: European Data Protection: In Good Health?, pp. 21--33.
  Springer (2012)

\bibitem{eckersley2011panopticlick}
Eckersley, P.: Panopticlick (2011)

\bibitem{getoor2012entity}
Getoor, L., Machanavajjhala, A.: Entity resolution: theory, practice \& open
  challenges. Proceedings of the VLDB Endowment  5(12),  2018--2019 (2012)

\bibitem{gomer2013network}
Gomer, R., Mendes~Rodrigues, E., Milic-Frayling, N., Schraefel, M.: Network
  analysis of third party tracking: User exposure to tracking cookies through
  search. In: Web Intelligence (WI) and Intelligent Agent Technologies (IAT),
  2013 IEEE/WIC/ACM International Joint Conferences on. vol.~1, pp. 549--556.
  IEEE (2013)

\bibitem{GoogleAnalytics}
Inc., G.: {Google Analytics} cookie usage on websites (2015),
  \url{https://developers.google.com/analytics/devguides/collection/analyticsjs/cookie-usage}

\bibitem{michael2013location}
Michael, K., Clarke, R.: Location and tracking of mobile devices:
  {\"U}berveillance stalks the streets. Computer Law \& Security Review  29(3),
   216--228 (2013)

\bibitem{mowery2011fingerprinting}
Mowery, K., Bogenreif, D., Yilek, S., Shacham, H.: Fingerprinting information
  in javascript implementations. Proceedings of W2SP  2 (2011)

\bibitem{Parra12TKDE}
Parra-Arnau, J., Perego, A., Ferrari, E., Forn{\'e}, J., Rebollo-Monedero, D.:
  Privacy-preserving enhanced collaborative tagging. {IEEE} Trans. Knowl. Data
  Eng.  26(1),  180--193 (Jan 2014),
  \url{http://dx.doi.org/10.1109/TKDE.2012.248}

\bibitem{Parra12DKE}
Parra-Arnau, J., Rebollo-Monedero, D., Forn{\'e}, J., Mu{\~n}oz, J.L., Esparza,
  O.: Optimal tag suppression for privacy protection in the semantic {W}eb.
  Data, Knowl. Eng.  81--82,  46--66 (Nov 2012),
  \url{http://dx.doi.org/10.1016/j.datak.2012.07.004}

\bibitem{rao2015they}
Rao, A., Schaub, F., Sadeh, N.: What do they know about me? contents and
  concerns of online behavioral profiles  (2015)

\bibitem{RAKE}
Rose, S., Engel, D., Cramer, N., Cowley, W.: Automatic keyword extraction from
  individual documents. Text Mining pp. 1--20 (2010)

\bibitem{veeningen2014line}
Veeningen, M., Piepoli, A., Zannone, N.: Are on-line personae really
  unlinkable? In: Data Privacy Management and Autonomous Spontaneous Security,
  pp. 369--379. Springer (2014)

\end{thebibliography}

\end{document}